\documentclass[twocolumn,aps,prl,showpacs,superscriptaddress,groupedaddress]{revtex4-1}
\usepackage[latin9]{inputenc}
\usepackage{geometry}
\usepackage{color}
\usepackage{amsmath}
\usepackage{amssymb}
\usepackage{graphicx}
\usepackage{esint}

\makeatletter

\usepackage{caption}

\@ifundefined{showcaptionsetup}{}{%
 \PassOptionsToPackage{caption=false}{subfig}}
\usepackage{subfig}
\makeatother

\begin{document}

\title{Nonequilibrium free energy estimation conditioned on measurement
outcomes}

\author{Shahaf Asban}

\affiliation{Faculty of Physics, Technion - Israel Institute of Technology, Haifa
32000, Israel}

\author{Saar Rahav}

\affiliation{Schulich Faculty of Chemistry, Technion - Israel Institute of Technology,
Haifa 32000, Israel}
\begin{abstract}
The Jarzynski equality is one of the most influential results in the
field of non equilibrium statistical mechanics. This celebrated equality
allows to calculate equilibrium free energy differences from work
distributions of nonequilibrium processes. In practice, such calculations
often suffer from poor convergence due to the need to sample rare
events. Here we examine if the inclusion of measurement and feedback
can improve the convergence of nonequilibrium free energy calculations.
A modified version of the Jarzynski equality in which realizations
with a given outcome are kept, while others are discarded, is used.
We find that discarding realizations with unwanted outcomes can result
in improved convergence compared to calculations based on the Jarzynski
equality. We argue that the observed improved convergence is closely
related to Bennett's acceptance ratio method, which was developed
without any reference to measurements or feedback. 
\end{abstract}

\date{\today}

\maketitle
\captionsetup{justification=raggedright,singlelinecheck=false}

\section{Introduction}

Information gained by measurements can be used to extract additional
work from a system. This deep connection between information and thermodynamics,
famously explored by Maxwell \cite{maxwell1871theory} and Szilard
\cite{Szilard1929}, is no longer merely a thought experiment. Several
experimental realizations of Maxwell's demon have been reported \cite{Toyabe2010,Berut2012,Koski2015},
and new theoretical insights have been gained \cite{Sagawa2010,Horowitz2010,Mandal2012,Barato2014,Parrondo2015}.
The renewed interest in this question is motivated by the development
of a new theoretical framework, stochastic thermodynamics \cite{Seifert2012},
that assigns thermodynamic interpretations to single realizations
of nonequilibrium processes.

One of the central results in the field is the Jarzynski equality
\cite{Jarzynski1997}. Consider a system that is prepared in thermal
equilibrium and then driven using an externally controlled variable
$\lambda\left(t\right)$ (for $0\le t\le t_{f}$). The Jarzynski equality,
\begin{equation}
\left\langle e^{-\beta w}\right\rangle =e^{-\beta\Delta{\cal F}},\label{eq:JE}
\end{equation}
connects the distribution of work values obtained from the nonequilibrium
process to the equilibrium free energy difference $\Delta{\cal F}={\cal F}_{\lambda\left(t_{f}\right)}-{\cal F}_{\lambda\left(0\right)}$
($\beta=1/k_{B}T$ is the inverse temperature.) Crooks has shown that
the Jarzynski equality follows from a detailed work relation that
compared the probability of time-reversed realizations \cite{Crooks1999},
while Hummer and Szabo have shown that it can be generalized for reaction
coordinate dependent free energies \cite{Hummer27032001}.

The Jarzynski equality can be used as a method of estimating free
energy differences in complex biomolecules, where it is hard \textcolor{black}{to}
maintain slow driving that keeps the system close to thermal equilibrium.
It tells us that one can still estimate equilibrium free energy differences
in far from equilibrium processes as long as the process is repeated
many times. Unfortunately, such an approach is not always practical
since the Jarzynski equality is known to suffer from a poor convergence
\cite{Gore2003,Jarzynski2006}. The ensemble average in Eq. (\ref{eq:JE})
may be dominated by rare events with exponentially large weights.
Unless such rare events are sampled sufficiently well, a free energy
estimate based on Eq. (\ref{eq:JE}) is likely to return very misleading
results. Motivated by potential applications in physics, chemistry
and computer simulations of biological molecules, several papers were
devoted to the convergence of the free energy calculations, and various
ways of improving it \cite{Jarzynski2006,Zuckerman2002,Gore2003,Harris2009,Kim2012,Shirts2003,Minh2008,Reid2010}.

The Jarzynski equality was generalized to include measurement and
feedback by Sagawa and Ueda \cite{Sagawa2010}. Their results were
later extended by Horowitz and Vaikuntanathan to processes with repeated
measurements \cite{Horowitz2010}. The role that feedback may play
in free energy calculations was not considered. It is only natural
to ask whether the inclusion of a Maxwell demon can improve the convergence
of free energy calculations. If so, {\em how} should one use the
information gained from the measurement?

In this paper we examine one possible way of employing measurements
in nonequilibrium free-energy calculations. The method we use is a
generalization of the Jarzynski equality that uses only realizations
with a given measurement outcome, while discarding the rest. We find
that the method can result in convergence rate that is faster than
that of calculations based on Eq. (\ref{eq:JE}). We then argue that
the mechanism behind this improved convergence is essentially the
one used in Bennett's acceptance ratio method \cite{Bennett1976}.
The latter is based on an efficient reweighing of realizations and
has no a priory relation to measurements and feedback.

The paper is organized as follows. In Sec. \ref{sec:expavg} we present
a method of calculating the free energy difference using only \textcolor{black}{the
realizations of a nonequilibrium process that were measured to have
a given} outcome. Sec. \ref{sec:conv} is devoted to a discussion
of the convergence of free energy calculations based on the method
presented in \ref{sec:expavg}. In Sec. \ref{sec:oscillator} we study
a simple process for which analytical expressions for the number of
realizations required for convergence can be found. We use these estimates
to argue that convergence can be accelerated by discarding realizations
and show the connections to Bennett's acceptance ratio method. In
Sec. \ref{sec:pulling} we study numerically a simple model for a
hairpin pulling experiment and show that much of the qualitative behavior
seen in the model of Sec. \ref{sec:oscillator} persists for more
complicated and realistic setups. We discuss the results in Sec. \ref{sec:dis}.

\section{Nonequilibrium free energy calculations conditioned on a measurement's
outcome}

\label{sec:expavg}

The Jarzynski equality (\ref{eq:JE}) holds for processes in which
a system is driven away from thermal equilibrium by \textcolor{black}{an
external variation of parameters according to a protocol $\lambda(t)$.
$\lambda(t)$ denotes a control parameter that enters into the system's
Hamiltonian ${\cal H}_{\lambda}(\vec{p},\vec{q})$. In this section
we present one way of utilizing measurements and feedback in nonequilibrium
free energy calculations. For this purpose, let us consider a similar
process in which the state of the system is measured at an intermediate
time, $t_{m}$, and an outcome $m$ is found (with probability $p_{m}$).
One can then apply feedback by modifying the driving to $\lambda_{m}(t)$
for $t>t_{m}$, in an outcome dependent way. (The same final value,
$\lambda_{m}(t_{f})=\lambda(t_{f})$, should be used in all protocols
to ensure that they all have the same final Hamiltonian.)}

Let us denote by $\gamma(t)\equiv\left\{ \vec{p}_{i}(t),\vec{q}_{i}(t)\right\} $
a realization of the process, encoding all the relevant information
regarding the systems evolution. This realization is a solution of
the time evolution equation, which can be either Hamiltonian or stochastic.
For every realization one can calculate thermodynamic quantities such
as heat and work characterizing the process. For the purpose of free-energy
estimation the work performed on the system $w(\gamma)=\int_{0}^{t_{f}}dt\dot{\lambda}\frac{\partial{\cal H}_{\lambda}(\gamma(t))}{\partial\lambda}$
is the most relevant.

One possible way of estimating the free-energy difference in processes
with measurement and feedback is based on a generalization of the
Jarzynski equality that require also an estimation of the information
gained by measurements in each realization \cite{Sagawa2010,Horowitz2010}.
Here we examine a different, but related approach, which has the advantage
of being valid for error-free measurements.

Let us now imagine that we preform a process with measurement and
feedback and collect all the work values corresponding to realizations
that were measured to be at a given outcome $m$. Can one calculate
$\Delta{\cal F}$ using the exponential average 
\begin{equation}
\left\langle e^{-\beta w}\right\rangle _{\vert m}\equiv\lim_{N_{m}\rightarrow\infty}\frac{1}{N_{m}}\underset{i=1}{\overset{N_{m}}{\sum}}e^{-\beta w(\gamma_{i})},\label{eq:expavg}
\end{equation}
computed from these work values? A short calculation, to be detailed
below, shows that $\Delta{\cal F}$ can be calculated from the realizations
with a given outcome by using 
\begin{equation}
\frac{p_{m}}{p_{m}^{R}}\left\langle e^{-\beta w}\right\rangle _{\vert m}=e^{-\beta\Delta{\cal F}},\label{eq:IFT-meas}
\end{equation}
where $p_{m}^{R}$ is the probability to satisfy $m$ at $\bar{t}_{m}=t_{f}-t_{m}$
in the reverse process. In this reverse process the system is prepared
in equilibrium with control parameter $\bar{\lambda}(0)=\lambda(t_{f})$.
It is then driven using the time-reversed protocol $\bar{\lambda}_{m}(t)\equiv\lambda_{m}(t_{f}-t)$.

Equation (\ref{eq:IFT-meas}) offers an interesting and different
way of estimating the free energy difference $\Delta{\cal F}$ from
an experiment or simulation with measurements. It applies for error-free
measurements. It also allows to obtain an independent estimate for
each outcome, offering the possibility that the calculation may converge
faster for one of the outcomes. We will see that this is indeed the
case in Secs. \ref{sec:oscillator} and \ref{sec:pulling}. Equation
(\ref{eq:IFT-meas}) was mentioned in passing in the beautiful experimental
realization of a Maxwell's demon reported by Toyabe, et. al. \cite{Toyabe2010},
and is closely related to earlier work regarding dissipation by Kawai,
Parrondo, and Van den Broeck \cite{kawai2007dissipation}. More recently,
Ashida {\em et. al.} pointed out that this equation allows to obtain
an achievable bound on the work that can be extracted from a process,
and offered qualitative guidelines on how to maximize it \cite{Ashida2014}.
The application of Eq. (\ref{eq:IFT-meas}) to free energy calculations
was not considered to the best of our knowledge.

The derivation of Eq. (\ref{eq:IFT-meas}) is fairly straightforward,
and was briefly described in \cite{Toyabe2010}. We give a detailed
derivation which also covers the case of several measurements below,
for completeness. The celebrated Crooks work relation \cite{Crooks1999}
states that 
\begin{equation}
\frac{{\cal P}_{\lambda}\left[\gamma\right]}{\bar{{\cal P}}_{\bar{\lambda}}\left[\bar{\gamma}\right]}=e^{\beta\left[w(\gamma)-\Delta{\cal F}\right]},\label{eq:Crooks}
\end{equation}
where $\bar{\gamma}$ is the time-reversed of $\gamma$, and ${\cal P}_{\lambda}[\gamma]$
and $\bar{{\cal P}}_{\bar{\lambda}}\left[\bar{\gamma}\right]$ denote
the probabilities of a realization and its time-reversed in the forward
and \textcolor{black}{reverse processes, respectively.}

Let us now add an error-free measurement to the process. In fact,
without any additional complication we can consider a set of $N$
measurements made at intermediate times $t_{1}\le t_{2}\le\cdots t_{N}$
in the forward process. We can denote the set of measured outcomes
in a single realization of the process by ${\cal M}\equiv\left\{ m_{1},m_{2},\cdots m_{N}\right\} $.
Assuming error-free measurements, the probability to measure these
specific outcomes in the process is given by the sum of probabilities
of all realizations that are consistent with the outcomes, 
\begin{equation}
p_{{\cal M}}=\sum_{\gamma\vert_{{\cal M}}}{\cal P}_{\lambda}\left[\gamma\right].\label{eq:defpm}
\end{equation}
Feedback can be applied by modifying the driving protocol following
each measurement.

Now we introduce a set of measurements that are performed in the reversed
process, so that they match the ones made in the forward process.
Specifically, the $i^{th}$ measurement in the reverse process is
made at time $t_{f}-t_{i}$, thereby reversing the order of measurements.
(To simplify the notations we assume that the measured quantities
are also even under time-reversal.) We note that the reverse process
has no feedback. The protocol used in the reverse process is chosen
in advance to be the time reversed of the protocol used in the forward
process with a given ${\cal M}$, and is not varied based on measurement
outcomes. The probability of a set of outcomes ${\cal M}^{\prime}=\left\{ m_{N}^{\prime},m_{N-1}^{\prime},\cdots,m_{1}^{\prime}\right\} $
in the reversed process is given by 
\begin{equation}
p_{{\cal M}^{\prime}}^{R}=\sum_{\gamma\vert_{{\cal M}^{\prime}}}\bar{{\cal P}}_{\bar{\lambda}}\left[\gamma\right].\label{eq:pmrev}
\end{equation}

Crucially, due to the error-free nature of the measurements, for each
realization $\gamma$ with outcomes ${\cal M}$ time-reversal matches
a realization $\bar{\gamma}$ of the reversed process with $\bar{{\cal M}}=\left\{ m_{N},m_{N-1},\cdots,m_{1}\right\} $.
The conditional average (\ref{eq:expavg}) is then given by 
\begin{equation}
\left\langle e^{-\beta w}\right\rangle _{\vert_{{\cal M}}}\equiv\sum_{\gamma\vert_{{\cal M}}}\frac{{\cal P}_{\lambda}\left[\gamma\right]}{p_{{\cal M}}}e^{-\beta w(\gamma)}.\label{eq:defcond}
\end{equation}
One can now use Crooks' re\textcolor{black}{lation (\ref{eq:Crooks})
to }obtain 
\begin{equation}
\left\langle e^{-\beta w}\right\rangle _{\vert_{{\cal M}}}=\sum_{\gamma\vert_{{\cal M}}}\frac{e^{-\beta\Delta{\cal F}}}{p_{{\cal M}}}\bar{{\cal P}}_{\bar{\lambda}}\left[\bar{\gamma}\right]=\frac{p_{\bar{{\cal M}}}^{R}}{p_{{\cal M}}}e^{-\beta\Delta{\cal F}}.\label{eq:usingcrooks}
\end{equation}
A simple rearrangement of terms give 
\begin{equation}
\frac{p_{{\cal M}}}{p_{\bar{{\cal M}}}^{R}}\left\langle e^{-\beta w}\right\rangle _{\vert_{{\cal M}}}=e^{-\beta\Delta{\cal F}}.\label{eq:eq3}
\end{equation}
Equation (\ref{eq:IFT-meas}) is obtained when one restricts the calculation
for a single measurement with outcome $m$.

\section{Convergence of nonequilibrium free energy calculations}

\label{sec:conv}

Equation (\ref{eq:IFT-meas}) is exact, and as a result when enough
realizations are sampled one can use it to accurately determine the
free energy difference. However, as is the case with free energy calculations
based on the Jarzynski equality (\ref{eq:JE}), calculations based
on a finite number of realizations may exhibit large errors due to
the need to sample rare events \cite{Jarzynski2006}.

Before discussing the convergence of nonequilibrium free energy estimation
based on Eq. (\ref{eq:IFT-meas}), let us briefly present some known
results regarding the convergence of free energy calculations in absence
of measurement and feedback. The estimate for the free energy difference
in this case is given by 
\begin{equation}
\Delta\hat{{\cal F}}(N)\equiv-k_{B}T\ln\left(\frac{1}{N}\sum_{i=1}^{N}e^{-\beta w_{i}}\right).\label{eq:JEestimate}
\end{equation}
The quality of this estimate is characterized by its systematic bias
\begin{equation}
{\cal E}(N)\equiv\left<\Delta\hat{{\cal F}}(N)\right>-\Delta{\cal F}\label{eq:JEbias}
\end{equation}
and its variance. Here $\left<\cdots\right>$ is an ensemble average
over many sets of $N$ work values. In the following we will mainly
focus on the systematic bias.

The convergence of free-energy calculations is known to qualitatively
depend on the dissipative work in the process. Specifically, ${\cal E}\left(1\right)=\left\langle w^{diss}\right\rangle =\left\langle w\right\rangle -\Delta{\cal F}$
determines the expected bias when only one work value is used. ${\cal E}\left(N\right)$
is a monotonically decreasing function of the number of trials $N$,
and therefore the dissipated work characterize the worst exponential
estimate \cite{Zuckerman2002,Gore2003}. For large enough $N$, when
the central limit theorem applies, ${\cal E}\left(N\right)\approx{\text{V}ar}\left[\exp\left(-\beta w^{diss}\right)\right]/2\beta N$
\cite{Zuckerman2002,Gore2003}. Finally, the number of trials needed
for convergence was estimated as $N^{*}\approx\exp\left(-\beta\left\langle w^{diss}\right\rangle ^{R}\right)$,
where $\left\langle w^{diss}\right\rangle ^{R}$ is the mean dissipated
work in the reverse process \cite{Jarzynski2006}.

We now turn to discuss free-energy estimation based on Eq. (\ref{eq:IFT-meas}).
One immediately notices a major difference compared to calculations
based on the Jarzynski equality (\ref{eq:JE}). Using Eq. (\ref{eq:IFT-meas})
to estimate $\Delta{\cal F}$ requires both the forward and reverse
process, due to the need to estimate the probability $p_{m}^{R}$.
Consider an attempt of estimating the free energy difference using
$N_{F}$ realizations of the forward process and $N_{R}$ realizations
of the reverse process. Imagine that in the forward {[}or the reverse{]}
process $N_{F}(m)$ {[}$N_{R}(m)$ respectively{]} of the realizations
were measured to be at $m$. As a result one estimates $p_{m}\simeq{N_{F}(m)}/{N_{F}}$
and $p_{m}^{R}\simeq N_{R}(m)/N_{R}$. The free energy difference
is estimated by 
\begin{multline}
\Delta\hat{{\cal F}}_{m}(N_{F},N_{R})=-k_{B}T\ln\left(\frac{1}{N_{F}}\sum_{i=1}^{N_{F}(m)}e^{-\beta w_{i}}\right)\\
+k_{B}T\ln\frac{N_{R}(m)}{N_{R}},
\end{multline}
where the sum over work values is restricted to realizations with
outcome $m$. The systematic bias of this estimate is obtained by
averaging over an ensemble of such processes, just as was done for
calculations based on the Jarzynski equality. In the following we
are interested in the qualitative behavior of this estimate, and it
will therefore suffice to discuss the case of $N_{F}=N_{R}=N/2$,
where $N$ is the total number of simulations used to obtain the estimate.

Obtaining a useful estimate of $\Delta{\cal F}$ requires one to accurately
calculate the two probabilities $p_{m},p_{m}^{R}$, as well as the
exponential average $\left<e^{-\beta w}\right>_{|m}$. Estimation
of the probabilities can be difficult when they are small. For example,
when $p_{m}^{R}\ll1$ one needs to repeat the reverse process of the
order of $N_{R,m}^{*}\sim\frac{1}{p_{m}^{R}}$ times to obtain a reasonable
estimate of the probability. When $p_{m}^{R}$ is very close to $1$
one has $N_{R,m}^{*}\sim1/(1-p_{m}^{R})$, but in this case $\Delta{\cal F}$
is less sensitive to errors in the estimation of $p_{m}^{R}$. The
number of realizations of the forward process $N_{F,m}^{*}$ needed
for a calculation of $p_{m}$ is determined similarly.

What is left is to estimate the number of realizations needed for
convergence of $\left<e^{-\beta w}\right>_{|m}$, under the assumption
that $p_{m},p_{m}^{R}$ are known. This exponential average behaves
just like its counterpart in the Jarzynski equality (\ref{eq:JE}),
except that only realizations with a specific outcome are used to
generate the work distributions of the forward and reverse processes.
These distributions of work values with an outcome are obtained from
the full work distribution by discarding realizations with the wrong
outcome. The resulting distribution must be divided by $p_{m}$ or
$p_{m}^{R}$ to be normalized properly. For error-free measurements
the Crooks relation can be used to map between a realization of the
forward process and its time-reversed counterpart in the reverse process.
Both these properties were used in the derivation of Eq. (\ref{eq:IFT-meas})
in Sec. \ref{sec:expavg}. In addition, application of the Jensen
inequality to Eq. (\ref{eq:IFT-meas}) leads to a second-law-like
inequality 
\begin{equation}
\left\langle \Sigma\right\rangle _{m}\equiv\left\langle w\right\rangle _{\vert m}-\Delta{\cal F}+k_{B}T\ln\frac{p_{m}^{R}}{p_{m}}\geq0.\label{eq:2nd law-meas}
\end{equation}
$\left\langle \Sigma\right\rangle _{m}$ depends on the measurement
outcome in two ways. Only work values of realizations with the outcome
$m$ are used to calculate the mean work. In addition, $\left\langle \Sigma\right\rangle _{m}$
depends on the probabilities $p_{m}$ and $p_{m}^{R}$. The last term
in (\ref{eq:2nd law-meas}) can be interpreted as the difference in
information gained by measuring $m$ in the forward process and in
its reversed counterpart. $\left\langle \Sigma\right\rangle _{m}$
is therefore related to both the dissipation in the process and the
information involved in restricting the realizations to a specific
outcome.

The exponential average in Eq. (\ref{eq:IFT-meas}) behaves like its
counterpart in the Jarzynski equality, just with renormalized work
distributions, and as a result with an exponent of $\Delta{\cal F}-k_{B}T\ln\frac{p_{m}^{R}}{p_{m}}$
instead of $\Delta{\cal F}$. This means that the previously known
estimates for the accuracy and convergence of the exponential average
can be modified to apply to $\left<e^{-\beta w}\right>_{|m}$, as
long as the dissipated work is replaced by $\left\langle \Sigma\right\rangle _{m}$.
Accordingly ${\cal E}(1)=\left<\Sigma\right>_{m}$ is the mean bias
of $\Delta\hat{{\cal F}}$ when a single realization with outcome
$m$ is used in the calculation of the exponential average (and $p_{m}$,
$p_{m}^{R}$ are known).

The bias ${\cal E}(1)$ serves as an upper bound for averages employing
more realizations of the forward process with this outcome. For large
enough $N_{F}(m)$, when the central limit theorem applies, ${\cal E}\left(N_{F}(m)\right)\approx{\text{V}ar}_{m}\left[\exp\left(-\beta\Sigma\right)\right]/2\beta N_{F}(m)$,
just like the behavior of the bias for calculations based on Eq. (\ref{eq:JE}),
see Refs. \cite{Zuckerman2002,Gore2003}. Finally, and most importantly,
the number of realizations needed for the convergence of $\left<e^{-\beta w}\right>_{|m}$
scales as 
\begin{equation}
N_{F,m}^{*,exp}\sim\frac{1}{p_{m}}\exp\left(\beta\left<\Sigma\right>_{m}^{R}\right),\label{eq:newexpest}
\end{equation}
where $\left<\Sigma\right>_{m}^{R}$ is calculated from the reverse
process. The factor of $1/p_{m}$ expresses the fact that only a fraction
of the realizations of the forward process will be measured to be
at $m$, and therefore be us\textcolor{black}{ed in }the calculation
of $\left<e^{-\beta w}\right>_{|m}$.

Collecting everything together we have 
\[
N_{m}^{*}\simeq N_{R,m}^{*}+\max\left\{ N_{F,m}^{*},N_{F,m}^{*,exp}\right\} 
\]
as a crude estimate for the total number of realizations required
for convergence. Note that, as in Ref. \cite{Jarzynski2006}, the
estimate in Eq. (\ref{eq:newexpest}) is obtained under the assumption
that the dominant realizations of the process are rare and hard to
sample. Under this assumption $N_{F,m}^{*,exp}$ is expected to be
considerably larger than $N_{F,m}^{*}$, and therefore 
\begin{equation}
N_{m}^{*}\simeq N_{R,m}^{*}+N_{F,m}^{*,exp}.\label{eq:finalest}
\end{equation}

Both the choice of the measurement and the application of feedback
can affect the convergence of free-energy calculations. We will explore
how this occurs using simple models in the rest of this paper. The
results will show that there is a tradeoff where improved convergence
of the exponential average may come at the expense of difficulty in
estimating one of the probabilities of measurement outcome. Nevertheless,
it will become clear that judicious choice of measurement and feedback
can indeed result in improved convergence.

\section{Measurements as a tool for improving convergence - a solvable model}

\label{sec:oscillator}

Gaining an intuitive understanding \textcolor{black}{of} the convergence
of free-energy calculations is not easy, since several important factors
play a simultaneous role. The convergence of non-equilibrium free
energy estimation based on the Jarzynski equality (\ref{eq:JE}) can
be improved by modifying the driving protocol to reduce dissipation.
The convergence of calculations based on Eq. (\ref{eq:IFT-meas})
\textcolor{black}{is} even more subtle, as it is affected by additional
factors. One is feedback, namely, the ability to tailor different
driving protocols to different measurement outcomes. Another is the
ability to estimate the free energy difference independently from
realization with different outcomes. The convergence need not be the
same for different outcomes, and in fact can be greatly improved by
chosing the measured quantity intelligently.

It is therefore desirable to study simple processes in which the different
factors affecting convergence can be separated. In this section we
investigate the convergence of a simple \textit{instantaneous} process.
Since the initial and final Hamiltonians are fixed, the only thing
that can affect the convergence is the choice of measurements, which
in turn determine which realizations are kept or discarded. The example
studied below will clarify how a judicious choice of measurement can
accelerate the convergence of nonequilibrium free energy calculations
based on Eq. (\ref{eq:IFT-meas}). It will become apparent that the
mechanism by which separation of realizations according to outcomes
improves convergence is closely related to Bennett's acceptance ratio
method \cite{Bennett1976}.

\subsection{The shifted harmonic oscillator}

Let us consider a process in which a particle is initially placed
in an harmonic potential $V(x)=\frac{1}{2}kx^{2}$, and is then allowed
to reach thermal equilibrium at temperature $T$. At time $t=0$ the
location of the particle is measured. There are many ways of bunching
together information about the particle location into several discrete
measurement outcomes. Here we use two coarse-grained outcomes: all
positions where $x\le a$ are said to be in region $I$, whereas as
$x>a$ is in region $II$. $a$ is a parameter characterizing the
division between the two outcomes.

\begin{figure}[h]
\noindent \begin{centering}
\includegraphics[bb=100bp 100bp 750bp 470bp,clip,scale=0.35]{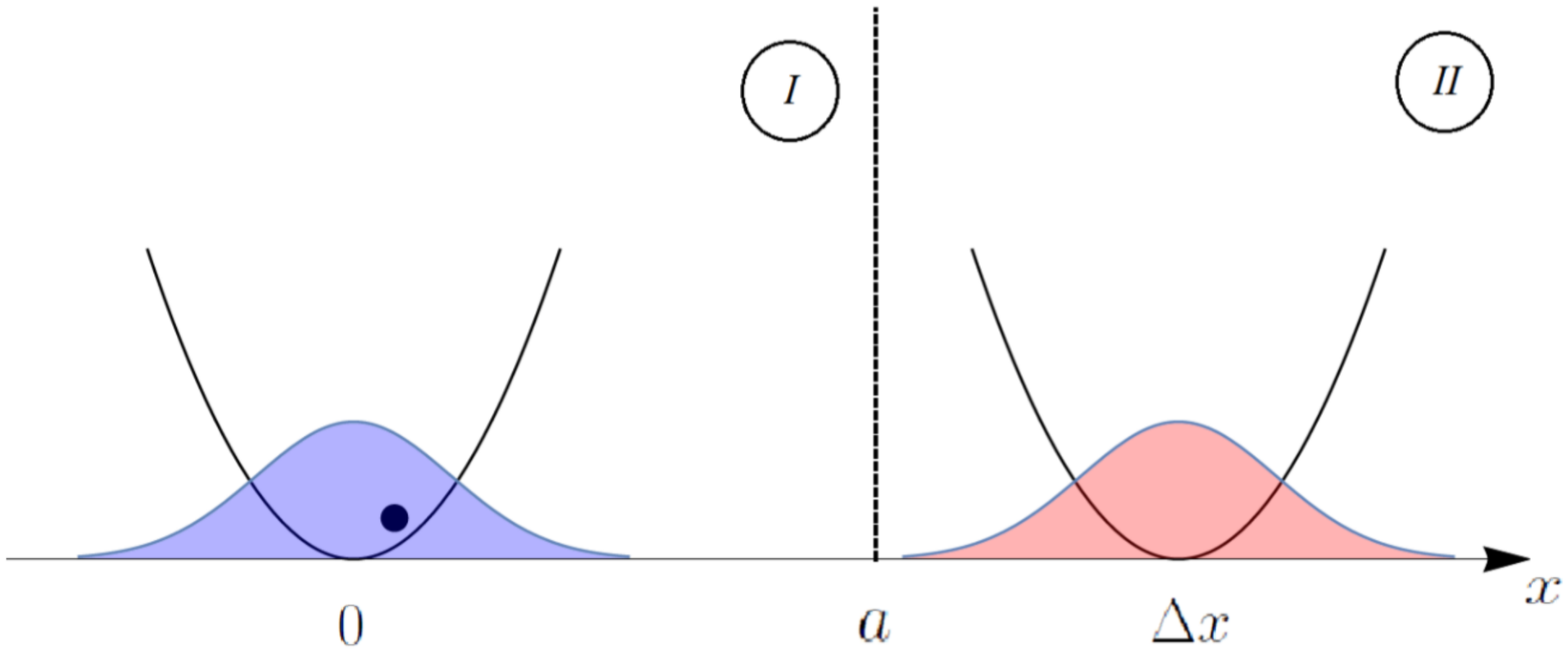} 
\par\end{centering}

\caption{Schematic depiction of the quenched harmonic potential. The initial
potential $V(x)$ is centered at $x=0$, whereas the potential after
the shift $V_{f}(x)$ is centered around $x=\Delta x$. The vertical
line divides between locations belonging to outcomes $I$ and $II$.
The filled curves depict the equilibrium probability distribution
of the potentials $V$ and $V_{f}$.\label{fig:harmonic}}
\end{figure}

Following the measurement the potential is suddenly shifted by a distance
$\Delta x$, so that it is given by $V_{f}(x)=\frac{1}{2}k(x-\Delta x)^{2}$.
The initial and final potential are depicted in Fig. \ref{fig:harmonic}.
It is clear that for this process $\Delta{\cal F}=0$. However, when
$\beta k\Delta x^{2}\gg1$ free energy calculations will suffer from
poor convergence, and many realizations will be needed in order to
obtain an accurate estimate for $\Delta{\cal F}$. The simple model
studied in this section allows to obtain an analytical estimate for
the number of realizations needed for convergence. 

Eq. (\ref{eq:IFT-meas}) also requires an investigation of a reverse
process. In this reverse process the system is initially equilibrated
in $V_{f}(x)$. The potential is then changed suddenly to $V(x)$,
and the particle location is measured.

The sudden nature of the process means that the system has no time
to evolve. All the information about a realization is given by its
initial condition, which is sampled out of the equilibrium distribution.
There is simply no time to apply feedback. Nevertheless, Eq. (\ref{eq:IFT-meas})
holds for realizations with either outcome, for any value of the parameter
$a$. We wish to know for what outcome and which value of $a$ the
convergence is fastest, and why.

\subsection{Estimation of the number of realization needed for convergence}

\label{sunseq:harmonicestimates}

Let us first give an estimate for the number of realizations needed
for convergence in free-energy calculations based on the Jarzynski
equality (\ref{eq:JE}). Here there is no measurement and only the
forward process is needed. The probability that the particle is initially
at $x$ is given by 
\begin{equation}
p_{F}(x)=\sqrt{\frac{\beta k}{2\pi}}e^{-\frac{1}{2}\beta kx^{2}}.\label{eq:inPx}
\end{equation}
When the harmonic potential is suddenly shifted work is done on the
particle. This work is given by 
\begin{equation}
w(x)=V_{f}(x)-V(x)=\frac{1}{2}k\Delta x^{2}-kx\Delta x.\label{eq:harmonicw}
\end{equation}

The so-called dominant realizations, which must be sampled to ensure
convergence, were identified by Jarzynski to be the time-reversed
of typical realizations of the reverse process \cite{Jarzynski2006}.
For the simple example considered here the initial equilibrium probability
distribution of the reverse process is 
\begin{equation}
p_{R}(x)=\sqrt{\frac{\beta k}{2\pi}}e^{-\frac{1}{2}\beta k(x-\Delta x)^{2}},\label{eq:inPx}
\end{equation}
and therefore the typical realizations in the reverse process are
those located near $x\simeq\Delta x$. For $\beta k\Delta x^{2}\gg1$
these realizations are in the far tail of the distribution $p_{F}(x)$,
and are exponentially unlikely. The number of realizations needed
for convergence therefore scales like 
\begin{equation}
N_{JE}^{*}\sim\frac{1}{p_{F}(\Delta x)}\sim e^{\frac{1}{2}\beta k\Delta x^{2}}.\label{eq:estJE}
\end{equation}
The symbol $\sim$ is used to mean that these expressions have the
same dominant exponential factor, but may differ by slower varying
prefactors that may, for instance, includes powers of $\beta k\Delta x^{2}$.
We employ such a crude asymptotic approximation since: i) we will
not be able to determine the correct prefactor when discussing convergence
with a measurement, and ii) we will consider parameter values such
that the exponentials are dominant, and knowledge of them will suffice
for comparison between alternative calculations.

We turn now to estimate the number of realizations required if one
wishes to estimate $\Delta F$ from Eq. (\ref{eq:IFT-meas}). It will
be instructive to consider calculations employing realizations with
outcomes $I$ and $II$ separately. One of the main difference between
calculations based on Eq. (\ref{eq:IFT-meas}) and on Eq. (\ref{eq:JE})
is that the former requires one to employ both the forward and reverse
process. Specifically, the combination $p_{m}\left<e^{-\beta w}\right>_{|m}$
is estimated by repeating the forward process, whereas $p_{m}^{R}$
is obtained from realizations of the reverse process. As discussed
in the previous section the costs of both estimates should be taken
into account.

Let us first focus on a calculation in which one estimates $\Delta F$
from realizations with outcome $II$. Let us also assume that the
parameter $a$ is located in the region between $0$ and $\Delta x$,
where $\beta ka^{2},\beta k(a-\Delta x)^{2}\gg1$. This means that
$a$ is in the tails of both $p_{F}(x)$ and $p_{R}(x)$, simplifying
many of the estimates in this section.

As discussed in the previous section one requires roughly 
\[
N_{F,II}^{*,exp}\sim\frac{1}{p_{II}}e^{\beta\left<\Sigma\right>_{II}^{R}}
\]
realizations of the forward process to estimate the exponential average
$\left<e^{-\beta w}\right>_{|II}$. The total cost of the calculation
with the $II$ outcome is therefore given by 
\begin{equation}
N_{II}^{*}(a)\sim\frac{1}{p_{II}}e^{\beta\left<\Sigma\right>_{II}^{R}}+\frac{1}{p_{II}^{R}}.
\end{equation}

Estimation of $N_{II}^{*}(a)$ is quite straightforward for $a$ values
that satisfy $\beta ka^{2},\beta k(a-\Delta x)^{2}\gg1$. $p_{II}^{R}\simeq1$
and therefore the difficulty in estimating $\Delta{\cal F}$ is solely
due to the forward process. In contrast $p_{II}\sim\exp\left(-\frac{1}{2}\beta ka^{2}\right)\ll1$
must be considered but this factor appears twice in the estimate.
Once as a prefactor, and once in $\left<\Sigma\right>_{|II}^{R}=\left<w\right>_{|II}^{R}+k_{B}T\ln\frac{p_{II}}{p_{II}^{R}}$
The two terms cancel each other. What is left is to estimate $\left<w\right>_{|II}^{R}$,
the mean work done by realizations of the reverse process with outcome
$II$. Since realizations with outcome $I$ are very rare in the reverse
process this amounts to estimating the mean work in this process,
for which typical realizations have $x\simeq\Delta x$. As a result
$\left<w\right>_{|II}^{R}\simeq V(\Delta x)-V_{f}(\Delta x)=\frac{1}{2}k\Delta x^{2}$.
Collecting everything, we find 
\begin{equation}
N_{II}^{*}(a)\sim e^{\frac{1}{2}\beta k\Delta x^{2}}\sim N_{JE}^{*}.\label{eq:estimateR}
\end{equation}

We have just found out that for a range of $a$ values the convergence
of calculations based on Eq. (\ref{eq:IFT-meas}) with outcome $II$
and of calculations based on the Jarzynski equality (\ref{eq:JE})
are equally challenging. Interestingly, convergence of $\left<e^{-\beta w}\right>_{|m}$
in Eq. (\ref{eq:IFT-meas}) is faster than that of $\left<e^{-\beta w}\right>$
in Eq. (\ref{eq:JE}), but this is compensated by the fact that most
realizations of the forward process will have the wrong outcome. Does
this mean that free energy calculation based on Eq. (\ref{eq:IFT-meas})
always exhibit the same convergence as calculations based on the Jarzynski
equality? No. As we shall shortly see, realizations with the outcome
$I$ exhibit very different behavior.

\textcolor{black}{The estimation of convergence for realizations with
outcome $I$ proceeds along similar lines. However, in this case $p_{I}=\int_{-\infty}^{a}p_{F}(x)dx\simeq1$,
while $p_{I}^{R}\sim e^{-\frac{1}{2}\beta k(\Delta x-a)^{2}}\ll1$.
In this case the dominant realizations that one must sample to obtain
a reasonable estimate for the free energy difference are the most
typical realizations of the reverse process {\em subject to the
constraint} $x\le a$. These are rare realizations, located in the
tail of the probability $p_{R}(x)$. In fact, the Gaussian shape of
$p_{R}(x)$ me}an that the most typical realizations of the reversed
process with the constraint are those with $x\simeq a$. For these
realizations the mean dissipated work is roughly 
\begin{equation}
\left<w\right>_{|I}^{R}\simeq V(a)-V_{f}(a)=ka\Delta x-\frac{1}{2}k\Delta x^{2}.
\end{equation}

The number of realizations required to estimate the combination $p_{I}\left<e^{-\beta w}\right>_{|I}$
is therefore 
\[
\frac{1}{p_{I}}e^{\beta\left<\Sigma\right>_{I}^{R}}\sim e^{\beta\left<w\right>_{|I}^{R}-\ln p_{I}^{R}}\sim e^{\frac{1}{2}\beta ka^{2}}.
\]
Since $p_{I}^{R}\ll1$ one also requires of the order of $1/p_{I}^{R}$
realizations of the reverse process. The total number of realizations
needed for a reasonable estimate of the free energy difference is
therefore 
\begin{equation}
N_{I}^{*}(a)\sim e^{\frac{1}{2}\beta ka^{2}}+e^{\frac{1}{2}\beta k(\Delta x-a)^{2}},\label{eq:costl}
\end{equation}
where the first term comes from the number of times the forward process
is performed while the second term comes from the number of times
one should realize the reverse process. Considering that $\beta ka^{2},\beta k(\Delta x-a)^{2}\gg1$
one immediately sees that $N_{I}^{*}(a)$ is minimal when $a=\Delta x/2$.
In that case $N_{I}^{*}(\Delta x/2)\sim e^{\frac{1}{8}\beta k\Delta x^{2}}\ll N_{JE}^{*}$.
Convergence of free energy estimation based on Eq. (\ref{eq:IFT-meas})
is therefore much faster in comparison to that of a calculation based
on Eq. (\ref{eq:JE}) \textcolor{black}{when} $a=\Delta x/2$ and
only realizations with outcome $I$ are used. Convergence is, in fact,
accelerated by ignoring realizations with the ``unwanted'' outcome
$II$!!

One can intuitively understand the reasons for different convergence
rates of calculations employed realizations with different outcomes
by looking at the dominant realizations which one needs to sample
in each case. For outcome $II$ the dominant realizations are those
with $x\simeq\Delta x$. These are also the dominant realizations
in calculations based on the Jarzynski equality (\ref{eq:JE}). These
realizations are very unlikely in the forward process, but are typical
for the reverse process.

The dominant realizations of the calculation with outcome $I$ are
located at $x$ values which are slightly smaller but close to $a$.
These are still rare realizations of the forward process, but sampling
them is much more likely than sampling realizations with $x\simeq\Delta x$.
To obtain the correct free energy difference from realization with
this outcome Eq. (\ref{eq:IFT-meas}) mandates that the reverse process
should also be used and $p_{I}^{R}$ estimated. One also needs realizations
with $x\simeq a$ for this purpose. The additional cost of sampling
such realizations is large, but also much less than the cost of sampling
forward realizations with $x\simeq\Delta x$. The improved convergence
is therefore achieved by designing the measurement (by choosing $a$)
in a way that modifies the dominant realizations of the process. A
good choice of $a$ results in a calculation with dominant realizations
that divide the numerical cost between the forward and reverse processes.
The nonlinear dependence of this cost on parameters results in a total
cost which is much lower than that of a naive calculation based on
Eq. (\ref{eq:JE}).

\subsection{Connection to Bennett's acceptance ratio method}

The estimation of the free energy difference based on Eq. (\ref{eq:JE}),
which was used as a benchmark in the previous subsection, is somewhat
naive. Better convergence can be attained by using the acceptance
ratio method proposed by Bennett in 1976 in the context of thermodynamic
perturbation theory \cite{Bennett1976}. While originally developed
for computations in which the Hamiltonian is changed in a sudden manner,
the method is applicable also for finite-time processes \cite{Shirts2003,Kim2012}.

The acceptance ratio method is based on the identity \cite{Shirts2003,Bennett1976}
\begin{equation}
e^{-\beta\Delta{\cal F}}=\frac{\left<f(w)\right>_{F}}{\left<f(-w)e^{-\beta w}\right>_{R}},\label{eq:FRavg}
\end{equation}
which holds for any function $f$. Bennett showed that the statistical
variance in estimates of $\Delta{\cal F}$ is minimal when the function
$f$ is chosen to be the Fermi function 
\begin{equation}
f(w)=\frac{1}{1+\frac{n_{F}}{n_{R}}e^{\beta(w-\Delta{\cal F})}}\label{eq:FF}
\end{equation}
where $n_{F}$ and $n_{R}$ are the number of realizations used to
sample the forward and reverse process respectively. Shirts \textit{et
al.} demonstrated that the acceptance ratio method emerges from maximum
likelihood considerations \cite{Shirts2003}.

To make a connection with the model studied here, for which we obtained
analytical estimates for number of realizations needed for convergence
in Sec. \ref{sunseq:harmonicestimates}, we note that the best convergence
was achieved when the numerical load was equally divided between the
forward and reverse processes. Substituting the Fermi function and
$n_{F}=n_{R}$ in Eq. (\ref{eq:FRavg}), we find 
\begin{equation}
e^{-\beta\Delta{\cal F}}=\frac{\left<\frac{1}{1+e^{\beta(\Delta{\cal F}-w)}}e^{-\beta w}\right>_{F}}{\left<\frac{1}{1+e^{\beta(\Delta{\cal F}+w)}}\right>_{R}}.\label{eq:Bennetteq}
\end{equation}
The Fermi functions in the numerator of Eq. (\ref{eq:Bennetteq})
give weights of order unity to realizations of the forward process
with $\beta w-\beta\Delta{\cal F}\gg1$. In contrast, the weights
are exponentially small when $\beta w-\beta\Delta{\cal F}\ll-1$.
There is a smooth transition between these two regions, which is centered
around $w=\Delta{\cal F}$.

For the shifted harmonic oscillator studied in Sec. \ref{sunseq:harmonicestimates}
$\Delta{\cal F}=0$. In addition, the sudden dynamics results in a
simple connection between the particle's location and the work done
during the process, see Eq. (\ref{eq:harmonicw}). $x=\Delta x/2$
is precisely the point where $w=\Delta F=0$. The region where $w>0$
is equivalent to $x<\Delta x/2$ and similarly $w<0$ for $x>\Delta x/2$.
The Fermi function in the numerator of Eq. (\ref{eq:Bennetteq}) therefore
gives larger weights to points that satisfy $x<\Delta x/2$, that
is, to points located in region $I$ ($a=\Delta x/2$). Since the
work done in the reverse process is minus the work done in the forward
process the Fermi function in the denominator of Eq. (\ref{eq:Bennetteq})
also favors points in $I$.

When the Fermi functions in Eq. (\ref{eq:Bennetteq}) are replaced
by step functions, either by hand, or by lowering the temperature
to 0, only realizations with $x<\Delta x/2$ are kept. Realizations
with $x>\Delta x/2$ are given vanishing weights and are effectively
discarded. Replacement of the Fermi function with a step function
therefore replaces Eq. (\ref{eq:Bennetteq}) with Eq. (\ref{eq:IFT-meas})
for the outcome $m=I$. This reveals that the improved convergence
found in Sec. \ref{sunseq:harmonicestimates} is closely connected
to Bennett's acceptance ratio method. Specifically, the nonequilibrium
free energy calculation for sudden processes, based on separating
realizations according to outcomes, is at best an approximation of
Bennett's acceptance ratio method, in which the only weights used
are $0$ or $1$.

\section{A model of a hairpin pulling experiment}

\label{sec:pulling}

The example studied in Sec. \ref{sec:oscillator} was designed to
be particularly simple to allow for a derivation of analytical estimates
for the number of realizations needed for convergence. This simplification
came with a cost. The model had instantaneous dynamics that left no
time for the application of feedback. But Eq. (\ref{eq:IFT-meas})
holds also for processes with feedback. Interestingly, inclusion of
feedback can either improve or hinder the convergence rate. On one
hand feedback can be used to make the process more reversible. On
the other hand, the time that it takes the system to respond to the
feedback is also the time in which the measurement loses some of its
predictive value regarding the work accumulated during the realization.

To see how this interplay between effects works in practice we study
numerically a model that mimics an experiment in which an RNA hairpin
is pulled open \cite{Ritort2002}. The pulling process occurs over
a finite time interval such that the system has time to respond to
changes in the pulling protocol following the measurement. The model
we study is chosen to exhibit the characteristic behavior of an actual
pulling experiment but is intentionally simplified. This makes it
easier to gather sufficient statistics, and more importantly to gain
qualitative understanding of its behavior.

\subsection{The two state model}

\begin{figure}[h]
\begin{centering}
\includegraphics[scale=0.4]{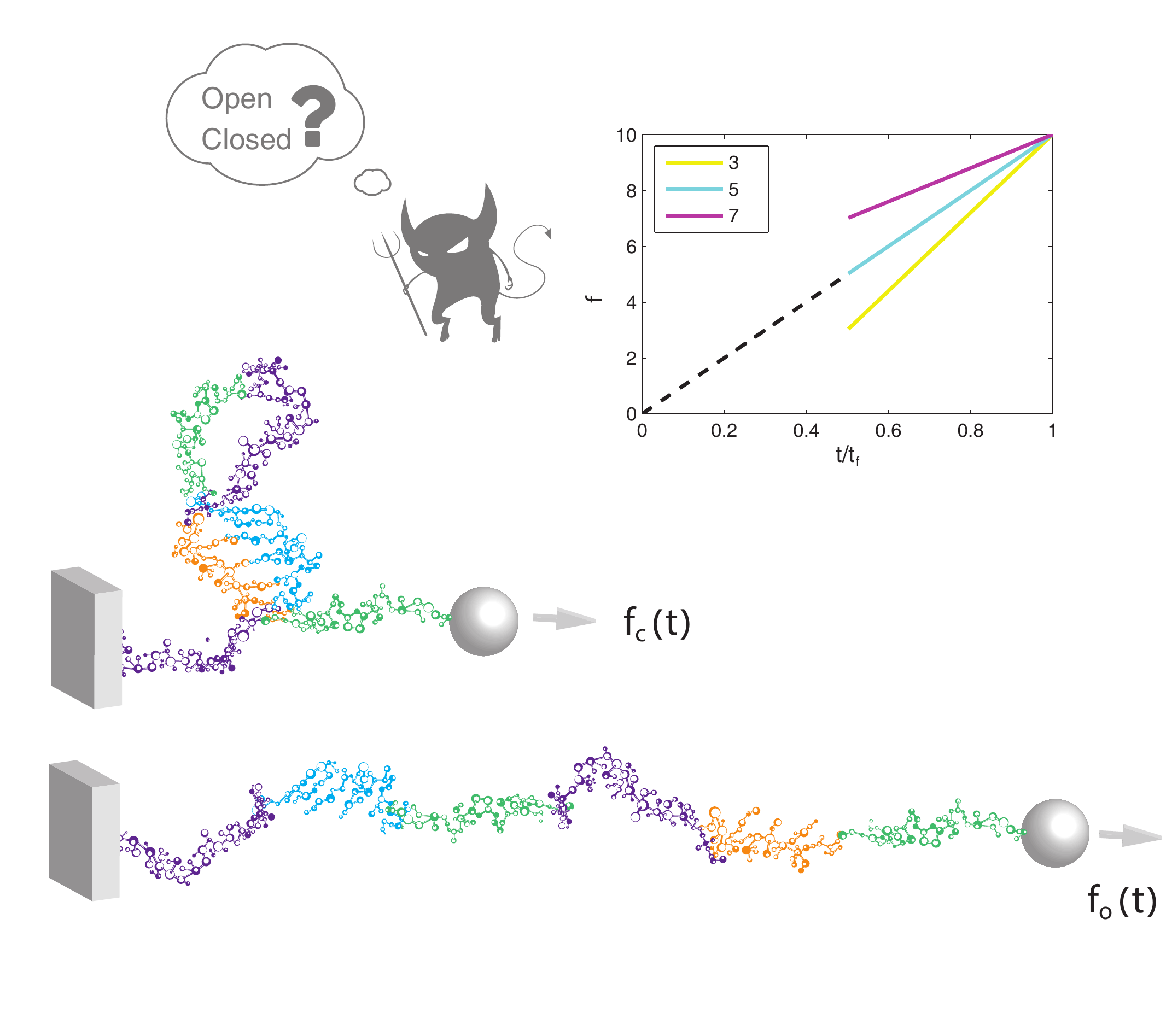} 
\par\end{centering}

\caption{Schematic depiction of a process where an RNA hairpin is pulled open
by an external force. The hairpins is modeled as a two state system,
and can either be in an open or a closed configuration. A measurement
is done in an intermediate time, here $t_{f}/2$, and feedback is
applied by changing the pulling force according to the outcome. This
change is restricted to a sudden jump in the value of the force, followed
by linear variation. Several possible force protocols are depicted
in the inset. \label{fig:Schematic-diagram}}
\end{figure}

Ritort, Bustamante, and Tinoco have used a simple model of a hairpin
pulling experiment to study the convergence rate of free energy calculations
\cite{Ritort2002}. Here we use the same model, but add measurement
and feedback to the process. This simple model of a hairpin has only
two states. A closed configuration with vanishing length and energy,
and an open state with length $l$ and energy $\Delta E_{0}-f\left(t\right)l$.
The force $f(t)$ is the external parameter used to drive the system
away from thermal equilibrium. The model is depicted qualitatively
in Fig. \ref{fig:Schematic-diagram}. Transitions between the states
are assumed to be Markovian, with rates $r_{c\rightarrow o}=1$ and
$r_{o\rightarrow c}=\exp\left[\Delta E-f\left(t\right)l\right]$,
where we used $\beta=k_{B}T=1$, $\Delta E=5$, and $l=1$. The subscripts
stand for 'open' and 'closed'. 

The pulling simulation starts from an equilibrium state, with $f\left(0\right)=0$,
where the system is very likely to be in the closed state. The force
is then increased according to a known protocol in order to open the
hairpin. When the system is in the open state work is done on it with
rate $\dot{w}=-\dot{f}(t)l$, whereas no work is done when the hairpin
is closed. The Jarzynski equality holds for this process, with $\Delta{\cal F}=\ln\left\{ \left[1+\exp\left(-\Delta E\right)\right]/\left[1+\exp\left(-\Delta E+f\left(t_{f}\right)l\right)\right]\right\} $.
For large final forces $\Delta{\cal F}\approx\Delta E-f\left(t_{f}\right)l$,
allowing to easily estimate the binding energy from $\Delta{\cal F}$.

A clear cut comparison of the rate of convergence of free energy calculations
with and without measurement and feedback requires knowledge of the
optimal driving protocol in each case. However, the optimal driving
protocol is not known for the process with feedback. To get a reasonable
comparison we restrict ourselves to a one parameter family of protocols.
One can then find the best driving protocol by an easy numerical search.
The family of driving protocols we chose to use consists of forces
that change linearly with time. The duration of the process was chosen
to be $t_{f}=1/2$. The initial and final values of the force are
always given by $f(0)=0$ and $f(t_{f})=10$.

In all the protocols we have used, the force was initially given by
$f(t)=10t/t_{f}$. When measurement and feedback are \textcolor{black}{employed},
the state of the system is measured at $t_{m}=t_{f}/2$. Immediately
after the measurement the force is changed to a value $f_{+}\equiv f(t_{m}^{+})$.
This value is the tunable parameter in the family of driving protocols.
At times satisfying $t>t_{m}$ the force is changed linearly until
it reaches a protocol independent final value $f(t_{f})=10$ at the
end of the process. Several force protocols from this one-parameter
family are depicted in the inset of Fig. \ref{fig:Schematic-diagram}.

The dynamics of the two state system was explored using a kinetic
monte-carlo simulation based on the algorithm developed in Ref. \cite{Prados1997}.
The protocol duration was chosen to be $t_{f}=1/2$. For this duration
trajectories typically make few transitions between the states, and
some never even make a single transition during the pulling process.
One therefore expects that free energy calculations will not converge
easily.

\subsection{Outcome-work correlations}

For the instantaneous process studied in Sec. \ref{sec:oscillator},
the measurement outcome was directly related to the work done in the
process. In processes with a finite duration, a measurement preformed
at a given instance cannot fully determine the work accumulated during
the whole process. However, one can expect that the open and closed
outcomes will be correlated with the work done in the process due
to the different work rate $\dot{w}$ in the two states. The usefulness
of the measurement in the context of free energy calculations is therefore
related to the degree of correlations between the outcome and work.
Our first goal was to investigate this correlation.

\begin{figure}[h]
\begin{centering}
\includegraphics[bb=120bp 230bp 500bp 550bp,scale=0.55]{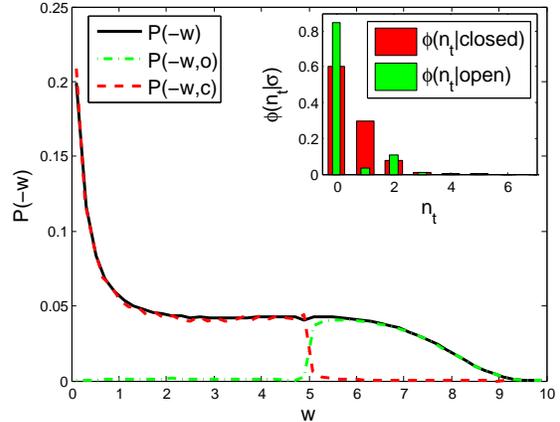} 
\par\end{centering}

\caption{\label{fig:PDF(w)}Correlations between work and measurement outcomes.
The solid line depicts the continuous part of the work distribution
for the linear protocol. The dashed red (grey) and green (light grey)
\textcolor{black}{lines} depict the joint distribution work and outcome.
For the selected parameters there is a clear correlation between the
work and measured outcome. Inset: the distribution of the number of
transitions made during the process, with realizations separated according
to the measurement outcome.}
\end{figure}

Fig. \ref{fig:PDF(w)} depicts the continuous part of the work distribution
obtained for the linear protocol (with $f_{+}=5$). We note that the
distribution also has a discrete part, at values $w=-10,0$, due to
trajectories that remain in the initial state throughout the process.
These are not shown. Performing a measurement allows to divide this
distribution into two pieces, depending on the outcome. They are also
depicted in Fig. \ref{fig:PDF(w)}. These marginal distributions of
work and outcome are clearly narrower than the full distribution.
The narrower work distribution associated with each outcome result
in faster convergence rate of the conditional exponential average
$\left<e^{-\beta w}\right>_{|m}$ compared to that of $\left<e^{-\beta w}\right>$
without a measurement. It should be clear that measurements that do
not result in narrower work distributions, will not be useful as an
avenue of accelerating convergence.

The neat division of the work distribution, depicted in Fig. \ref{fig:PDF(w)},
results from the fact that work accumulation is strongly coupled to
the state of the system and the relatively short duration of the process.
Most of the realizations contributing to the distribution depicted
in the figure make a single transition between the states. When this
transition precedes the meas\textcolor{black}{urement the outcome
will be open and the accumulated work will satisfy $w<-5$, and vice
versa. The inset depicts the distribution of the number of transitions
made in the process, conditioned on the measurement outcome. Nearly
$60\%$ of the trajectories starting in the closed state remain there
throughout the evolution, highlighting the fact that the system is
driven far from equilibrium.}

This explanation for the correlations between the measurement outcome
and the work done in the process, suggests that there is a tradeoff
between the usefulness of measurements and feedback. One expects that
the correlations between the outcome and work will be partially erased
in slower processes, where typical realizations make many transitions.
On the other hand, a longer duration gives the system more time to
respond to changes in the force, and thus increases the effectiveness
of feedback. For processes with very short duration the situation
is reversed. One expect better correlations between outcome and work
for judiciously designed measurements. But if the duration of the
process is short compared to the typical time between transitions,
the system barely reacts to changes in the force. This tradeoff between
the utility of measurements and feedback makes it difficult to use
feedback to accelerate the convergence rate beyond the improvement
due to the inclusion of measurements that was studied in Sec. \ref{sec:oscillator}.

\subsection{Convergence estimates}

The model of the pulling process can be used to test some of the qualitative
estimates for convergence discussed in Sec. \ref{sec:conv} in the
context of a finite time process. We start by assuming that the probabilities
$p_{m}$ and $p_{m}^{R}$ are known. In this case the rate of convergence
is determined by the convergence of the exponential average $\left<e^{-\beta w}\right>_{|m}$.
Does the convergence of this exponential average behaves like that
of calculations based on the Jarzynski equality, with $\left<\Sigma\right>_{m}$
playing the role of the dissipated work?

\begin{figure}[h]
\centering \includegraphics[bb=120bp 230bp 500bp 550bp,scale=0.57]{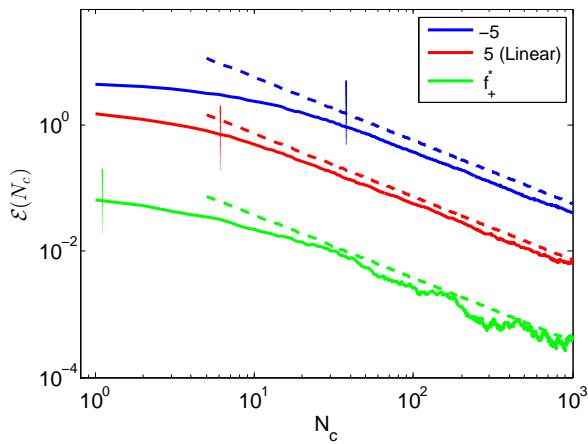}
\caption{The mean bias ${\cal E}\left(N_{c}\right)$, calculated under the
assumption that $p_{c}$ and $p_{c}^{R}$ are known, as a function
of the number of realizations with $m=closed$. The dashed lines correspond
to the predicted bias for large values of $N_{c}$, where the central
theorem applies. The vertical lines depict the estimated number of
realizations needed for convergence of $\left<e^{-\beta w}\right>_{|c}$.
The blue (upper) line corresponds to $f_{+,c}=-5$, the red (middle\textcolor{black}{)
line to the perfectly linear protocol, and the green (lower) line
to the protocol with $f_{+,c}=f^{*}$, namely to the protocol with
the lowest bias.}}
\label{fig:convmeasures} 
\end{figure}

Figure \ref{fig:convmeasures} depicts the mean bias ${\cal E}\left(N_{c}\right)$
for three different protocols. Since the purpose of the calculation
was to examine the convergence rate of the exponential average the
calculation was done using the exact values of $p_{c}$ and $p_{c}^{R}$.
One of the protocols we used was the completely linear protocol, with
$f_{+}=5$ and $\left<\Sigma\right>_{c}\simeq1.5$. We also studied
a protocol that is expected to have poor convergence rate, with $f_{+}=-5$
and $\left<\Sigma\right>_{c}\simeq4.5$. Finally, we took a protocol
with $f_{+,c}^{*}=10$, which (nearly) minimizes $\left<\Sigma\right>_{c}$,
with $\left<\Sigma\right>_{c}\simeq0.07$. Based on the discussion
of Sec. \ref{sec:conv} we expect that simulations using this protocol
would converge faster than its counterparts.

The results depicted in Fig. \ref{fig:convmeasures} indeed show that
this optimized protocol performs better \textcolor{black}{in comparison
to }its counterparts for all values of $N_{c}$. We see no indication
that the bias curves in Fig. \ref{fig:convmeasures} cross each other.
This, combined with the monotonic decrease of ${\cal E}(N)$ with
$N$ suggests that ${\cal E}(1)=\left<\Sigma\right>_{c}$ can be used
to \textcolor{black}{parameterize} the convergence of the exponential
average. Lower values of $\left<\Sigma\right>_{c}$ lead to more accurate
determination of $\Delta F$ at the same cost. The dashed lines correspond
to ${\cal E}\left(N_{c}\right)\approx{\text{V}ar}_{c}\left[\exp\left(-\beta\Sigma\right)\right]/2\beta N_{c}$,
which describes the large $N$ regime, where convergence is assured.
Finally, the horizontal lines mark the value $N_{c}^{*}=\exp\left(\beta\left\langle \Sigma\right\rangle _{c}^{R}\right)$,
suggested as a rough estimate for the number of realizations needed
for convergence \cite{Jarzynski2006}. One can see that this estimate
performs reasonably well for the linear and $f_{+}=-5$ protocols,
since it points to the region where the calculated bias (for large
$N_{c}$) starts to deviate from the numerically computed bias. In
contrast, this estimate for $N_{c}^{*}$ fails for the optimal protocol,
but this is expected since in this case $\left<\Sigma\right>_{c}<1$,
and the assumptions under which this estimate was derived are violated.

The numerical results depicted in Fig. \ref{fig:convmeasures} therefore
behave as predicted in the discussion in Sec. \ref{sec:conv}. This
supports the notion that the convergence of exponential averages over
realizations with a specific measurement outcome behave just like
the exponential average in the Jarzynski equality, as long as the
role of the dissipated work is played by $\Sigma$.

In many circumstances the probabilities $p_{m}$ and $p_{m}^{R}$
are not known and one must estimate them using realizations of both
the forward and reverse process, accordingly. Indeed, in our discussion
in Sec. \ref{sec:conv} we included the expected cost of this estimation,
see Eq. (\ref{eq:finalest}). For the simple example studied in Sec.
\ref{sec:oscillator} we have found a tradeoff between the number
of realizations needed to accurately estimate the relevant quantities
from the forward and reverse processes. When one became more efficient,
its counterpart became more troublesome. One may wonder whether a
similar tradeoff plays a role also for finite time processes that
includes feedback.

\begin{figure}[h]
\begin{centering}
\subfloat{\textcolor{black}{\centering{}\includegraphics[bb=105bp 240bp 500bp 560bp,clip,scale=0.57]{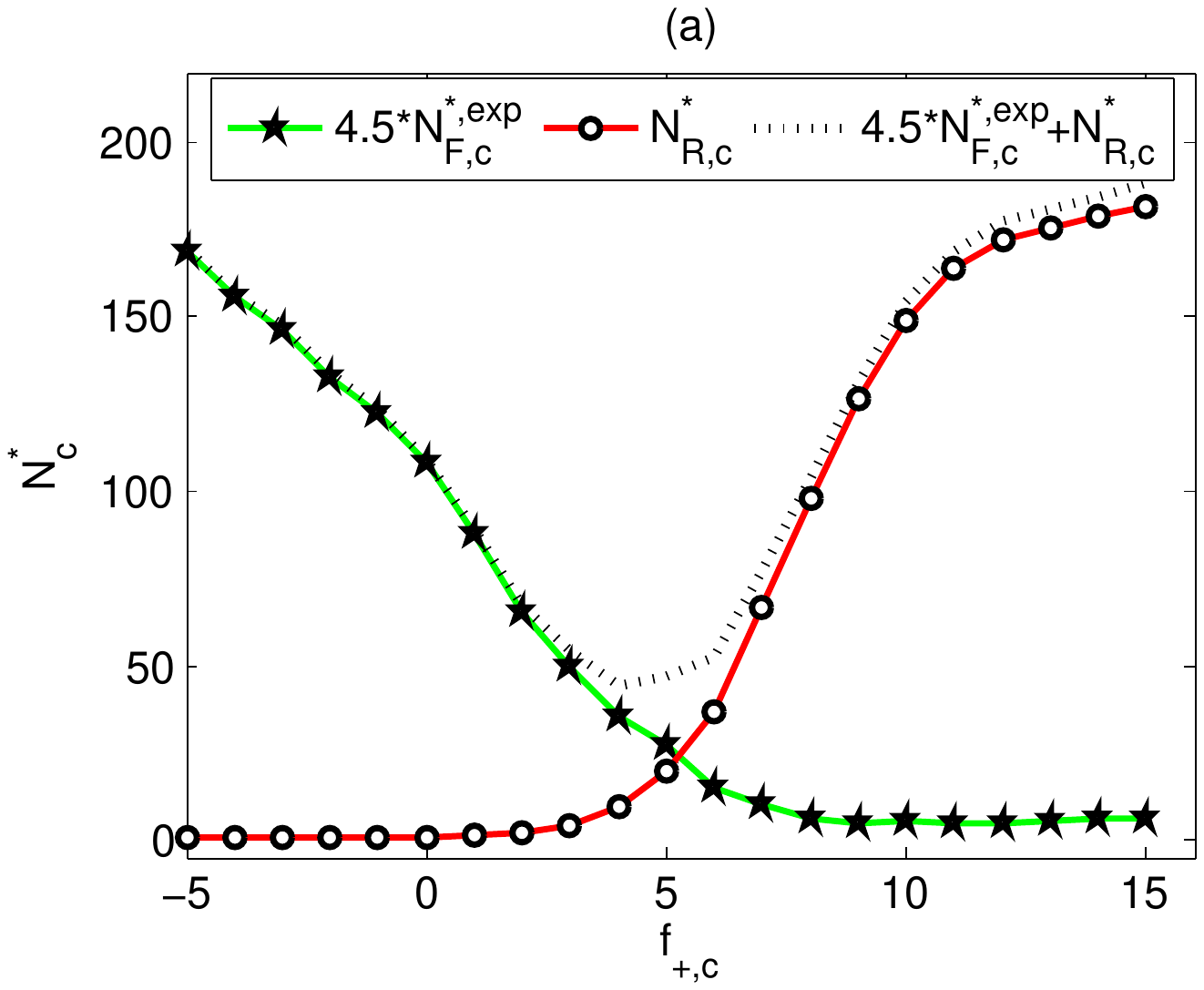} }}
\par\end{centering}

\begin{centering}
\subfloat{\textcolor{black}{\centering{}\includegraphics[bb=105bp 240bp 500bp 560bp,clip,scale=0.57]{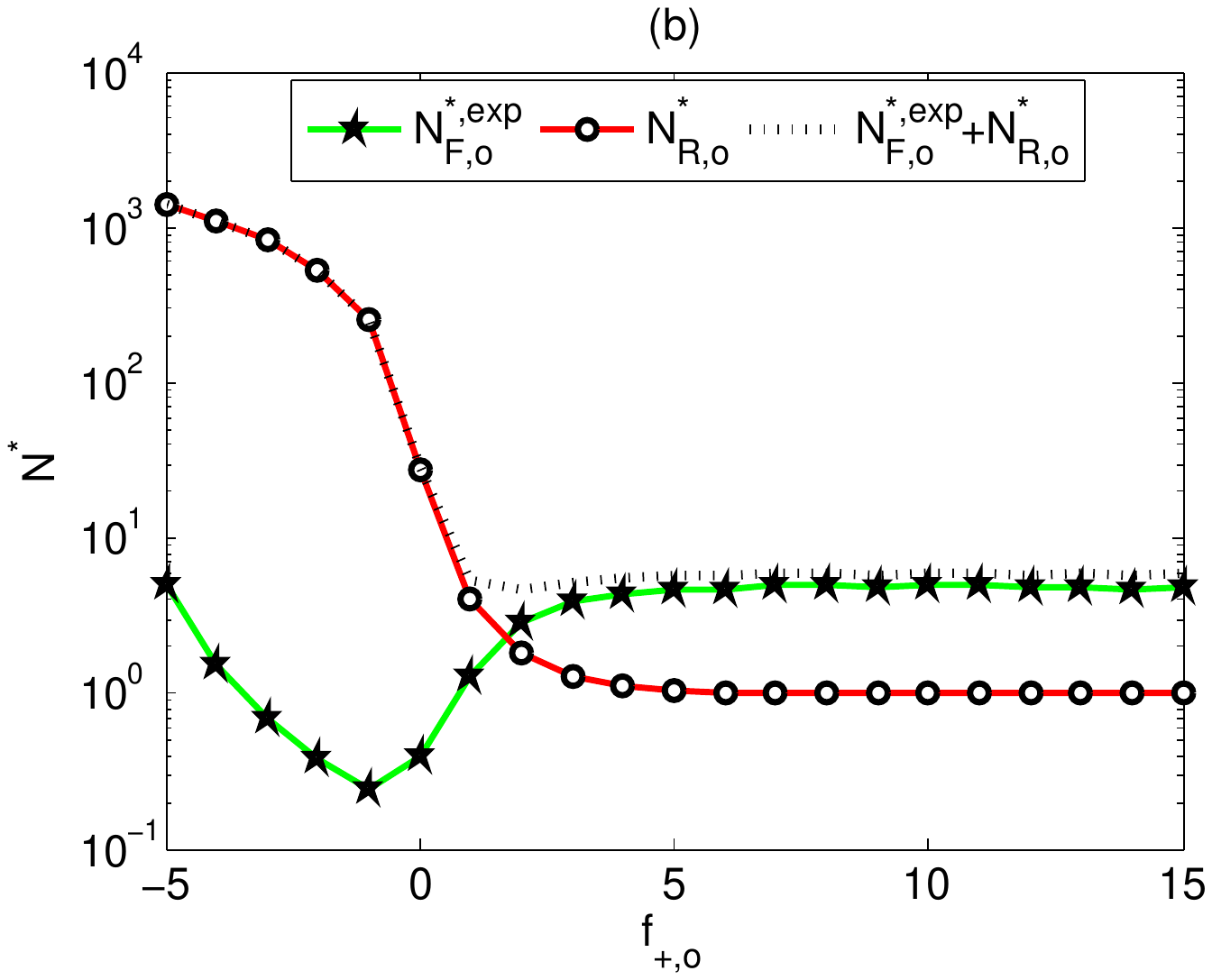} }}
\par\end{centering}

\textcolor{black}{\caption{\label{fig:Tradeoff}Estimates of the number of realizations required
for convergence of free energy calculations as a function of \textcolor{black}{$f_{+,m}$.
The solid-green line with the 5-pointed star symbols is proportional
to the number of realizations of the forward} process. (See discussion
in the main text.) The solid-red line with the 'o' symbols corresponds
to the number of realizations of the reversed process. The dotted
line is their sum. The top panel depicts results for realizations
with the closed outcome, while the bottom panel includes the estimates
for the open outcome.}
} 
\end{figure}

Fig. \ref{fig:Tradeoff} depicts results for $N_{F,m}^{*,exp}$ and
$N_{R,m}^{*}$ as a function of the feedback parameter $f_{+}$. Results
with the closed outcome are depicted in the top panel, whereas results
with the open outcome are included in the bottom panel. The results
in the top panel exhibit a clear tradeoff, in which $N_{F,c}^{*,exp}$
decrease with increasing $f_{+}$, while $N_{R,c}^{*}$ increases.

Let us recall that in Sec. \ref{sec:oscillator} realizations with
outcome $II$ exhibited a perfect tradeoff in costs, while for realizations
with outcome $I$ it was possible to find values of $a$ that divided
the cost between the forward and reversed trajectory in a way that
reduced the overall numerical cost. Trying to determine which of these
types of tradeoffs exist in the current model is tricky. The reason
is that we do not know the pre exponential factor of the estimates,
and the parameters used in the simulations are not extreme enough
to ensure that these pre-exponential factors are irrelevant. Reaching
qualitative insights, we chose the numerical prefactor of $N_{F,m}^{*,exp}$
in a way that attempts to create a perfect tradeoff. In the case of
the top panel of \ref{fig:Tradeoff} the sum of costs from the forward
and reverse trajectories was chosen to have the same value at the
edges of the range of $f_{+}$ values we used. With this somewhat
arbitrary choice we expect to find a flat curve in the case of a perfect
tradeoff. Instead the curve exhibits a minimum, suggesting that improved
convergence is possible in analogy with the results for outcome $I$
in Sec. \ref{sec:oscillator}.

The bottom panel of Fig. \ref{fig:Tradeoff} depicts the estimated
number of realizations needed for convergence for a calculation based
on the open outcome. For $f_{+,o}>3$ one notices that $N_{F,o}^{*,exp}$
is much larger than $N_{R,o}^{*}$. Moreover, both are essentially
constant in this region of parameters. This behavior is essentially
identical to the one found for realizations with outcome $II$ in
Sec. \ref{sec:oscillator}. The qualitative explanation for this behavior
is similar as well. The dominant realizations in a calculation based
on Eq. (\ref{eq:JE}) will be open at the measurement time. These
are the typical realizations in the reverse process. Variation of
$f_{+,o}$ in this range has limited ability to change this due to
the relatively short duration of the process. As a result the mean
work performed in the reverse process does not vary much, and $N_{F,o}^{*,exp}$
is nearly constant.

The behavior is quite different for $f_{+,o}<0$. In this region the
probability $p_{o}^{R}$ decreases sharply and as a result the numerical
cost of estimating $\Delta{\cal F}$ starts to be dominated by the
difficulty to sample the reverse process. The mean work of open trajectories
also varies considerably in this region. The decrease of $p_{o}^{R}$
reveals the reason for this behavior. For negative enough values of
$f_{+,o}$ typical realizations of the reverse process are no longer
open. In this regime feedback is strong enough to modify the dominant
realizations needed for an accurate estimate.

Overall, the results presented in Fig. \ref{fig:Tradeoff} show qualitative
behavior that is remarkably similar to the behavior of the shifted
harmonic oscillator studied in Sec. \ref{sec:oscillator}, despite
the obvious differences between the two setups. The fact that the
measurement outcome do not fully predict the value of the work does
not destroy this similarity due to the correlations depicted in Fig.
\ref{fig:PDF(w)}. The ability of feedback to modify the dynamics
is mostly seen in extreme values of the parameters, specifically for
$f_{+,o}<0$.

\subsection{Accuracy of free energy calculations}

\begin{figure}
\centering{} \includegraphics[bb=120bp 230bp 500bp 550bp,scale=0.6]{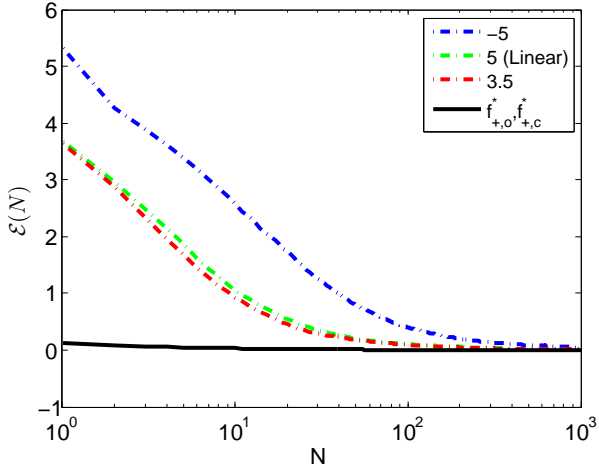}
\caption{Comparison of the mean bias with and without measurement and feedback.
The th\textcolor{black}{ree dashed-dotted lines depict the mean bias
of three calculations based on Eq. (\ref{eq:JE}). Three protocols
were used: The green line (second from top) corresponds to the completely
linear protocol, while the blue (top) line depicts a protocol with
$f_{+}=-5$. The red line (third from top) depicts the results for
a protocol optimized for minimal dissip}ated work $f_{+}=3.5$. The
solid-line corresponds for the bias from a calculation with feedback
and using Eq. (\ref{eq:IFT-meas}). For this curve the force protocols
were chosen to minimize $\left<\Sigma\right>_{c,o}$. The calculation
was performed under the assumption the values of the probabilities
$p_{c,o}$ and $p_{c,o}^{R}$ are known.\label{fig:calcwithp}}
\end{figure}

Fig. \ref{fig:calcwithp} highlights the drastic improvement that
inclusion of measurement and feedback can achieve if the probabilities
$p_{m}$ and $p_{m}^{R}$ are known in advance and do not have to
be estimated. The dashed-dotted lines correspond to free-energy calculations
based on the Jarzynski equality (\ref{eq:JE}) using several different
protocols. The solid line correspond to a calculation based on Eq.
(\ref{eq:IFT-meas}). It uses results from both measurement outcomes.
The protocols of both outcomes were optimized to minimize $\left<\Sigma\right>_{c,o}$,
with $f_{+,o}^{*}=8$ and $f_{+,c}^{*}=10$.

When the probabilities are not known, the results presented in Sec.
\ref{sec:oscillator} suggest that accelerated convergence is possible
if one chooses a measurement, outcome, and protocol that divide the
difficulty of the calculation between the forward and reverse process.
The results depicted in Fig. \ref{fig:Tradeoff} imply that the closed
outcome is the outcome for which such a division is possible.

\begin{figure}
\begin{centering}
\includegraphics[bb=110bp 220bp 500bp 550bp,clip,scale=0.6]{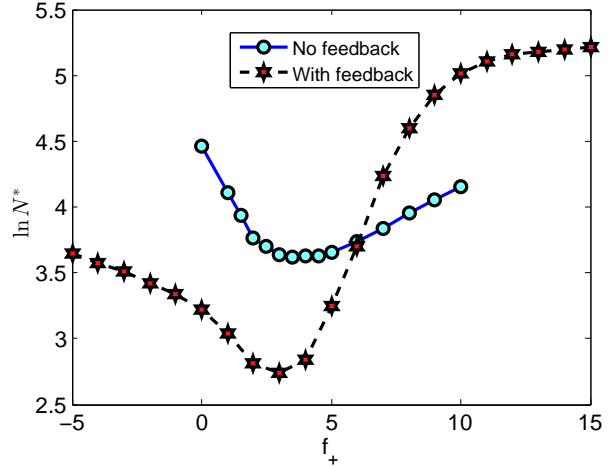} 
\par\end{centering}

\caption{Estimated number of realizations needed for convergence as a function
of the jump in the force value immediately after the measurement.
The start correspond to the estimated numerical cost of a calculations
based on Eq. (\ref{eq:IFT-meas}) with $m=closed$. The circles depict
the estimated cost of a calculation based on the Jarzynski equality
(\ref{eq:JE}). \label{fig:Estimated-number}}
\end{figure}

Fig. \ref{fig:Estimated-number} compares the estimated number of
realizations needed for convergence of calculations based on Eqs.
(\ref{eq:JE}) and (\ref{eq:IFT-meas}) respectively. The estimate
for the calculation with measurement and feedback is composed of the
two contributions presented in the top panel of Fig. \ref{fig:Tradeoff},
although they are summed without the prefactor that we included by
hand there.

The two estimates shown in Fig. \ref{fig:Estimated-number} allow
to choose protocols that are expected to exhibit the fastest rate
of convergence out of all the protocols in the family. For the calculation
based on the Jarzynski equality one should therefore choose $f_{+}\simeq3$.
Similarly, for the calculation that is based on Eq. (\ref{eq:IFT-meas})
$f_{+,c}\simeq3.5$ leads to the minimal number of realizations needed
for convergence. A comparison of both minima suggests that the calculation
which is based on Eq. (\ref{eq:IFT-meas}) should be the more accurate
of the two.

\begin{figure}[h]
\includegraphics[bb=105bp 230bp 500bp 600bp,clip,scale=0.57]{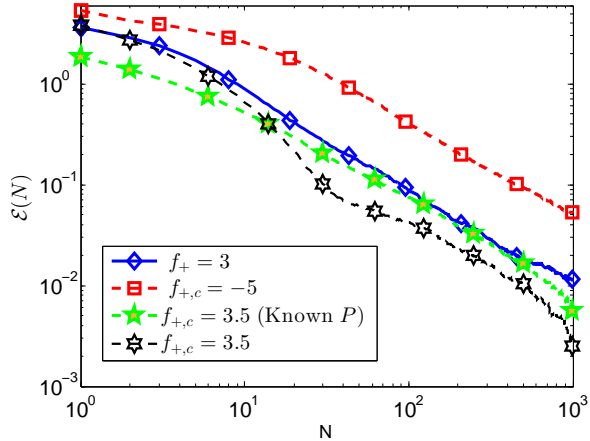}

\raggedright{}\caption{\label{fig:overall improvement} Comparison of the mean bias of the
free energy in calculations based on Eq. (\ref{eq:JE}) and Eq. (\ref{eq:IFT-meas}).
The solid line with the diamond symbols correspond to a calculation
based on Eq. (\ref{eq:JE}), with the fastest converging protocol
($f_{+}=3$). The dashed lines all correspond to calculations based
on Eq. (\ref{eq:IFT-meas}), where only realizations with the closed
outcome were used. The red (upper) line with square symbols depict
the mean bias for a bad choice of a protocol (with $f_{+,c}-5$).
Both the green (lowest) and black curves are obtained for the protocol
with $f_{+,c}=3.5$, which is expected be the most accurate based
on the estimate depicted in Fig. \ref{fig:Estimated-number}. The
\textcolor{black}{green line with 5-pointed star symbols depicts the
bias under the assumption that $p_{c}$ and $p_{c}^{R}$ are known.
The black curve with 6-pointed star symbols includes also the bias
due to the need to estimate these probabilities. }}
\end{figure}

Fig. \ref{fig:overall improvement} compares the mean bias of free
energy calculations with several different protocols. 

The blue-solid curve with the diamond symbols correspond to a calculation
based on the Jarzynski equality (\ref{eq:JE}). The value of $f_{+}=3$
was chosen to minimize the dissipative work in the reverse process.
This curve therefore serves as a baseline to which calculations based
on Eq. (\ref{eq:IFT-meas}) are compared. We note that the bias for
the completely linear protocol ($f_{+}=5$) was also calculated, and
found to lie almost on top of this curve. The results for the linear
protocol were omitted to avoid cluttering the figure.

The dashed lines in Fig. \ref{fig:overall improvement} correspond
to the bias for several calculations based on Eq. (\ref{eq:IFT-meas}).
The results of Figs. \ref{fig:Tradeoff} and \ref{fig:Estimated-number}
suggest that it is best to use realizations with the closed outcome,
and that the smallest number of realizations required for convergence
is obtained for $f_{+,c}\simeq3.5$. The c\textcolor{black}{urve with
$f_{+,c}=-5$ is therefore expected to show comparatively large errors.
The results depicted in Fig. \ref{fig:overall improvement} indeed
show that fo}r this protocol the mean bias is consistently worse than
that of the calculation based on Eq. (\ref{eq:JE}). The two other
curves correspond to calculations done with a protocol that has $f_{+,c}=3.5$.
They are much more accurate as expected.

\textcolor{black}{The black line with 6-pointed stars corresponds
to the bias from a calculation that takes into account the need to
estimate the probabilities $p_{c}$ and $p_{c}^{R}$. In contrast,
the green line with star symbols depicts the bias obtained under the
assumption that these probabilities are known. It is presented here
to highlight the differences between the two contributions to the
bias. Interestingly, for most of the range of $N$ values in the figure
the two biases have opposing signs and therefore they partially cancel
each other.}

\textcolor{black}{A prominent feature of all the curves in Fig. \ref{fig:overall improvement}
is that they all become parallel to each other for large values of
$N$. This expresses the fact that all the biases scale as $C/N$
there. Such scaling is expected due to the central limit theorem,
see e.g. the results depicted in Fig. \ref{fig:convmeasures}.}

\textcolor{black}{The results depicted in Fig. \ref{fig:overall improvement}
show that the bias of the calculations based on Eq. (\ref{eq:IFT-meas})
{[}with $f_{+,c}\simeq3.5${]} is smaller than the bias of a calculation
based on Eq. (\ref{eq:JE}). However, the difference between the biases
is quite modest. Here it is crucial to note that the results depicted
in Fig. \ref{fig:overall improvement} were obtained in a parameter
regime which is not in the asymptotic region where convergence is
exceedingly difficult. Indeed, one sees from Fig. \ref{fig:Estimated-number}
that $N_{c}^{*}\simeq15$ and $N^{*}\simeq35$. Moreover, these estimates
neglected the role of an unknown prefactor which may be become relevant
for not too high values of $N_{c}^{*}$, $N^{*}$. In light of this
the modest difference between the two calculations is not unexpected.
Based on the qualitative considerations made in Secs. \ref{sec:conv}
and \ref{sec:oscillator} we expect a much more pronounced difference
between the two methods if the parameters are tuned such that $N_{c}^{*}$
and $N^{*}$ are larger by two or more orders of magnitude. Unfortunately,
such a numerical investigation will require exponentially more computer
resources.}

\section{Discussion}

\label{sec:dis}

In this manuscript we have investigated a method of calculating equilibrium
free energy differences using repetitions of a nonequilibrium process
with measurements and feedback. The method is based on Eq. (\ref{eq:IFT-meas}),
which allows to calculate $\Delta F$ from realizations with a specific
measurement outcome. We have found that improved rate of convergence
can be achieved by keeping realizations with one outcome and discarding
the rest. While the rate of convergence obtained with this method
can be superior to that of a {\em naive calculation} using the
Jarzynski equality, our results suggest that it may be very difficult
to design setups in which the calculation will perform better than
calculations employing Bennett's acceptance ratio method. This conclusion
is based on the observation that the method studied here exhibits
built in tradeoffs where improved convergence in one part of the calculation
makes the estimation of other parts more challenging.

The first of these tradeoffs is related to the way that the separation
into different outcomes affect the number of realizations needed to
be sampled in the forward and reverse process. The shifted harmonic
oscillator studied in Sec. \ref{sec:oscillator} clearly exhibits
this tradeoff. Free energy calculations based on Eq. (\ref{eq:IFT-meas})
require both the forward and reverse processes, just like calculations
based on Bennett's acceptance ratio method. The forward process is
used to estimate $p_{m}\left<e^{-\beta w}\right>_{|m}$, while the
reverse process is needed to estimate the probability $p_{m}^{R}$.
In the case of outcome $II$ ($x>a$), changing $a$ could be used
to ease the calculation of the exponential average $\left<e^{-\beta w}\right>_{|II}$,
but that came at the expense of a lower value of $p_{II}$. The end
result was that one requires the same number of forward realizations.
For the other outcome, $x<a$, we found that improved convergence
of the forward process came at the expense of slower convergence rate
in its reversed counterpart. In this case one could choose the parameter
$a$ such that the numerical effort will be divided equally between
processes. As discussed in Sec. \ref{sec:oscillator} this mechanism
of accelerating the convergence rate is based on reweighing of realizations,
similar to the mechanism of Bennett's acceptance ratio method. Crucially,
Bennett's method is designed so that the weights are chosen for minimal
variance and maximal statistical likelihood of the estimator, and
are therefore superior to the weights used in Eq. (\ref{eq:IFT-meas}).

Interestingly, the results presented in Sec. \ref{sec:pulling}, obtained
for a finite time process with feedback, exhibit qualitatively similar
behavior. Fig. \ref{fig:Tradeoff} shows estimates for the number
of realizations needed for convergence of the forward and reverse
process. In this case the measurement is fixed, and the parameter
which is varied modifies the system's dynamics. There is no a priori
reason to expect behavior that is similar to the one encountered in
Sec. \ref{sec:oscillator}. Nevertheless, the results presented in
the top panel Fig. \ref{fig:Tradeoff} (for $m=closed$) exhibit qualitatively
similar behavior to that of realizations with outcome $I$ in Sec.
\ref{sec:oscillator}. Similarly, the results depicted in the bottom
panel of Fig. \ref{fig:Tradeoff} (for $m=open$) are qualitatively
similar to those found for $x>a$ in Sec. \ref{sec:oscillator}, at
least for $f_{+}>0$. Crucially, the dominant trajectories needed
for convergence of calculation based on Eq. (\ref{eq:JE}) are ones
with $m=open$ and $x>a$ in both setups. This suggests that\textcolor{black}{{}
the qualitative similarity of the tradeoffs is not accidental. It
is rather related to whether the trajectories that are conditioned
to a certain outcome include the set of dominant trajectories of the
calculation based on Eq. (\ref{eq:JE}), or not. In the latter case,
the restriction to a specific outcome results i}n a new group of dominant
trajectories that are consistent with the outcome.

The second tradeoff is between the usefulness of the weights assigned
to realizations and the feedback. Keeping only realizations with a
specific outcome is a way of reweighing realizations that uses only
the weights $0$ and $1$. From the considerations leading to the
Bennett's acceptance ratio method, a good choice of the weights should
depend on the work done in each realization. In the current method
the weights are based on the outcome rather than on the work, relying
on the strong coupling between the two at short time-scales. For process
of short duration the outcome and work can be correlated, suggesting
that one can design a measurement that would in principle result in
weights that are similar to those of the acceptance ratio method.
But short duration also means lack of time to respond to feedback.
For processes with long duration the situation is reversed. The dynamics
has ample time to respond to changes in parameters, but this comes
at the cost of reduced correlations between the outcome and work.
This hurts the efficiency of assigning outcome-based weights to realizations.

The results collected in Secs. \ref{sec:oscillator} and \ref{sec:pulling}
suggest that nonequilibrium free energy calculations based on Eq.
(\ref{eq:IFT-meas}) can show increased convergence rate compared
to naive calculations based on the Jarzynski equality (\ref{eq:JE}).
However, they are unlikely to perform better than calculations employing
Bennett's acceptance ratio method in repetitions of a single experiment.
The underlying reason is that the method uses the weights $0$ and
$1$, whereas the acceptance ratio method employees a continuous set
of weights, that are chosen to result in the maximal likelihood estimator,
which also exhibits the minimal variance. An opposite conclusion may
appear considering setups in which the probabilities $p_{m}$ and
$p_{m}^{R}$ are known from other sources. For instance if one can
solve a master or Fokker-Planck equation for the evolution of the
probability distribution, or alternatively run an experiment over
many copies of the system at the same time. If the probabilities are
known one can design the driving protocol to minimize $\left<\Sigma\right>_{m}$,
thereby reducing the difficulty of performing the exponential average.
In such cases one should expect drastic improvement in the convergence
rate, as is depicted in Fig. \ref{fig:calcwithp}.

Finally, the qualitative understanding regarding the convergence developed
here suggests a more promising approach for the inclusion of measurement
and feedback in nonequilibrium free-energy calculations. It is desirable
to use optimally chosen weights for any choice of feedback. This may
be possible for in the case of measurements with finite errors. A
version of the fluctuation theorem holds for such process \cite{Sagawa2010,Horowitz2010},
\[
\left<e^{-\beta w-I}\right>=e^{-\beta\Delta F}.
\]
This suggests that one can modify the arguments leading to Bennett's
acceptance ratio so they would hold for this pair of forward and reverse
process. The resulting weights would be a function of $\beta w+I$
instead of $\beta w$. The crucial point is that this is a good choice
of weights irrespectively of the driving protocol. One can then design
feedback that makes the process as reversible as possible. Horowitz
and Parrondo \cite{Horowitz2011} demonstrated that in certain cases
processes with measurement and feedback can be fully reversible.

Such reversibility is likely to be out of reach for processes whose
purpose is to forcibly change the system configuration in a finite
time, such as the pulling process of Sec. \ref{sec:pulling}. Nevertheless,
it will be of great interest to find out by how much one can reduce
the dissipation of a process by the addition of an optimally chosen
measurement and feedback, and by how much such a procedure accelerates
the convergence rate of free energy calculations. The investigation
of this \textcolor{black}{alternative} approach is left for future
work. 
\begin{acknowledgments}
We would like to thank Noa Marco-Asban for the graphical illustrations.
This work was supported by the the U.S.-Israel Binational Science
Foundation (Grant No. 2014405), by the Israel Science Foundation (Grant
No. 1526/15), and by the Henri Gutwirth Fund for the Promotion of
Research at the Technion 
\end{acknowledgments}

 \bibliographystyle{apsrev4-1}
\bibliography{Bibliography}

\end{document}